\documentclass[preprint,12pt]{elsarticle}

\usepackage[T1]{fontenc}
\usepackage[utf8]{inputenc}
\usepackage{amsmath}
\usepackage{booktabs}
\usepackage{graphicx}
\usepackage{adjustbox}
\usepackage{array}
\usepackage{multirow}
\usepackage[table]{xcolor}
\usepackage{hyperref}
\usepackage{tikz}
\usetikzlibrary{arrows.meta,positioning}

\journal{arXiv}
\date{}
\biboptions{authoryear}

\newcommand{\phigh}{P(\text{High})}

\begin{document}

\begin{frontmatter}

\title{Semi-Automatic Quantification of Bayesian Networks for Software Decision Support: Comparing WSA and RNM in a Software R\&D Organization}

\author[ise]{Mirko Perkusich\corref{cor1}}
\ead{mirko@virtus.ufcg.edu.br}
\author[ufcg]{Jo\~ao Nunes}
\ead{joao.nunes@copin.ufcg.edu.br}
\author[aarhus]{Emilia Mendes}
\ead{eme@ece.au.dk}
\author[ise]{Emanuel Dantas}
\ead{emanuel.dantas@virtus.ufcg.edu.br}
\author[ise]{Ademar Sousa}
\ead{ademar.sousa@virtus.ufcg.edu.br}
\author[ise]{Danyllo Albuquerque}
\ead{danyllo.albuquerque@virtus.ufcg.edu.br}
\author[ufcg]{Kyller C. Gorg\^onio}
\ead{kyller@dee.ufcg.edu.br}
\author[ufcg]{Angelo Perkusich}
\ead{perkusic@dee.ufcg.edu.br}

\cortext[cor1]{Corresponding author}

\address[ise]{Intelligent Software Engineering Group, VIRTUS, Federal University of Campina Grande, Rua Apr\'igio Veloso, 882, Bairro Universit\'ario, 58429-900 Campina Grande, Para\'iba, Brazil}
\address[ufcg]{Federal University of Campina Grande, Rua Apr\'igio Veloso, 882, Bairro Universit\'ario, 58429-900 Campina Grande, Para\'iba, Brazil}
\address[aarhus]{Department of Electrical and Computer Engineering, Aarhus University, Finlandsgade 22, 8200 Aarhus N, Denmark}

\begin{abstract}
Expert-driven Bayesian networks can support recurring software decisions when historical data are limited, but quantifying their conditional probability tables (CPTs) requires many probability judgments. Semi-automatic methods reduce direct elicitation, yet practitioners have little comparative evidence from operational decision processes. We report an embedded case study comparing the Weighted Sum Algorithm (WSA) and Ranked Nodes Method (RNM) as complete elicitation-and-quantification pipelines in two contexts within a single software research and development (R\&D) organization: feature selection for Internet of Things projects and user interface design selection. Within each context, the pipelines shared the graph, root priors, and decision evidence. We evaluated them through 15 expert-defined model-walkthrough scenarios and retrospective reconstructions of alternatives recorded in decision meetings. WSA matched 7/7 and 6/8 walkthrough expectations, whereas RNM matched 4/7 and 4/8. Both pipelines placed the selected alternatives near the top of the retrospective rankings. Their complete distributions nevertheless differed: the median total variation distance was 0.304 across four distinct feature patterns and 0.340 across seven distinct design patterns, rising to 0.890 for one design pattern. These results concern two ordinal value-estimation models in one organization. They show that similar shortlist behavior does not imply equivalent representations of uncertainty. For comparable software decision-support models, method selection and validation should consider node semantics, the judgments experts can provide, calibration requirements, complete output distributions, and the intended downstream use of the probabilities.
\end{abstract}


\end{frontmatter}

\section{Introduction}
\label{sec:introduction}

Software teams make recurrent decisions about features, designs, and technical alternatives under incomplete information and competing stakeholder concerns. Bayesian networks (BNs) can combine available observations with expert knowledge and make the assumptions behind such decisions explicit~\citep{pearl2014probabilistic,misirli2014bayesian,manzano2020method}. This capability is particularly useful when organization-specific variables matter and historical data are sparse. Building an expert-driven BN, however, requires both defining the graph and quantifying its conditional probability tables (CPTs). The latter can determine whether an otherwise useful decision-support model is feasible to construct and maintain.

Direct elicitation becomes demanding because CPT size grows exponentially with the number and cardinality of a node's parents. Repeated probability judgments can also introduce fatigue, inconsistency, and avoidable cognitive burden~\citep{druzdel2000building,das04generating,baker2010evaluating}. Semi-automatic methods reduce the input requested from experts and use a stated procedure to complete the remaining probabilities. The Weighted Sum Algorithm (WSA) and the Ranked Nodes Method (RNM) are two established alternatives, but they encode relationships differently and require different expert judgments and calibration choices~\citep{baker2009towards,laitila2016improving,laitila2018theoretical}.

Evidence that a method reduces the number of elicited values is not sufficient for organizational adoption. Practitioners also need to know whether the resulting model preserves expected behavior, supports the decisions for which it was built, and represents uncertainty in a way that is suitable for downstream use. A comprehensive review shows that alternative assumptions and information sources used to derive CPTs can affect BN results and identifies comparative evaluation and sensitivity analysis as continuing needs~\citep{rohmer2020uncertainties}. These questions are difficult to answer from methodological examples alone. Existing comparisons of WSA and RNM are limited, and operational evidence in software decision processes is especially scarce. Recent research likewise identifies calibration, traceability, and expert burden as continuing concerns in BN quantification~\citep{laitila2022advancing,podofillini2023traceable,blomaard2025burden}.

We address this practice-oriented gap through an embedded case study in one software research and development (R\&D) organization. The two embedded contexts concern feature selection for forthcoming sprints and user-interface design selection. They provide organization-specific Bayesian value-estimation models, judgments from professionals involved in the decisions, and alternatives recorded during decision meetings. Within each context, WSA and RNM used the same graph, root-node priors, and decision evidence, so the comparison focuses on the implemented elicitation and quantification pipelines. We ask:

\begin{description}
    \item[RQ1:] How often do WSA- and RNM-quantified models agree with the predefined modal output in hypothetical walkthrough scenarios?
    \item[RQ2:] How do the rankings and probability distributions produced by WSA and RNM correspond to stakeholder selections recorded in historical decision meetings?
\end{description}

The walkthrough labels are expert-defined expectations rather than independently observed outcomes, and the historical choices are decisions rather than objective measurements of value. We therefore distinguish \emph{scenario agreement} from \emph{retrospective selection concordance}.

Earlier studies used the WSA-quantified models from these decision contexts to investigate value-based software engineering and the VALUE framework (described in Section~\ref{subsec:value})~\citep{mendes2018using,mendes2019using}. The present study reuses the corresponding VALUE records as common inputs, but addresses a different question: the RNM quantifications collected for the same contexts were not reported in those studies, nor were WSA and RNM compared as paired quantification pipelines. The present study makes three contributions. First, it reports how WSA and RNM were operationalized in two recurring software decision processes, including the elicitation roles, model structures, and retained RNM parameterizations. Second, it provides a paired comparison under fixed within-context conditions using both model-walkthrough expectations and recorded organizational decisions. Third, it derives practice-oriented criteria for selecting and validating a quantification pipeline, supported by a paired distribution-level analysis using total variation distance (TVD) to show when similar modes or rankings conceal materially different probability allocations. The evidence is bounded by the two models and one organization, while the evaluation procedure and lessons are relevant to comparable expert-driven software decision-support models.

The remainder of this article explains both techniques and related evidence (Section~\ref{sec:background}), the case-study design and analysis (Section~\ref{sec:method}), the results (Section~\ref{sec:results}), their implications (Section~\ref{sec:discussion}), and the limitations that bound the claims (Section~\ref{sec:threats}).

\section{Background and related work}
\label{sec:background}

Expert-driven BN quantification involves both obtaining conditional probabilities and assessing the consequences of the resulting model. The comprehensive review by \citet{rohmer2020uncertainties} provides a secondary synthesis of CPT derivation, uncertainty propagation, and sensitivity analysis. Following this distinction, this section introduces expert-driven quantification and the main families of semi-automatic CPT methods, describes WSA and RNM, and reviews empirical comparisons and complementary approaches to model evaluation.

\subsection{Expert-driven BN quantification}

A discrete BN represents variables as nodes and conditional dependencies as directed arcs. The graph and its local probability distributions together factorize the joint distribution, so evidence entered at one or more nodes updates beliefs elsewhere in the network~\citep{pearl2014probabilistic}. For a non-root node $Y$ with parents $X_1,\ldots,X_p$, a CPT specifies $P(Y\mid X_1,\ldots,X_p)$. If $Y$ has $r$ states and its parents have $q_1,\ldots,q_p$ states, the table has $r\prod_i q_i$ entries, or $(r-1)\prod_i q_i$ free probabilities once each row is normalized. As parents are added, or their cardinality grows, the table size grows fast, and so does the elicitation burden.

When representative data exist, CPTs can be estimated from observations. Expert-driven BNs become necessary when data are scarce, when variables capture organization-specific judgments, or when a model has to be built before enough observations accumulate. Eliciting every CPT row by hand is costly in that setting and tends to produce inconsistent judgments~\citep{druzdel2000building,baker2010evaluating}. Semi-automatic quantification reduces the burden on experts and derives the rest of the table using a stated model. Nothing about that is free: the resulting table depends on how the method represents causal interaction, ordinality, parent influence, and uncertainty.

\subsection{Families of semi-automatic CPT methods}

Rohmer's synthesis distinguishes elicitation of probabilities, simplification of causal relationships, and methods that fill CPTs from a reduced set of judgments~\citep{rohmer2020uncertainties}. These approaches reduce elicitation demands in different ways: by changing the questions asked, restricting the relationships represented, or deriving unassessed configurations from a smaller set of inputs.

Independence-of-causal-influence models are among the best-known reduced-elicitation approaches. Noisy-OR and Noisy-MAX replace a full table with parameters describing the separate influence of each cause~\citep{noisyor,noisymax,pearl2014probabilistic}. They work well when their assumptions fit the domain, but they commonly require distinct absent or baseline states, as well as conditional independence among the causal mechanisms. These restrictions can be a poor fit when several value factors interact, or when all states represent meaningful degrees of an ordinal construct~\citep{fenton2007using,baker2009towards}.

Interpolation methods select specific CPT rows and estimate the remaining rows from relationships among the assessed configurations~\citep{cain2001planning,tang2007developing,wisse2008relieving,podofillini2014aggregating}. Functional and piecewise interpolation can represent a broader range of relationships than a fixed canonical model, though the number and placement of anchors affect both the burden and fidelity. Other approaches derive CPTs from weighted functions~\citep{das04generating,fenton2007using,roed2009use}, or replace direct probability judgments with pairwise comparisons~\citep{monti2000dealing,chin2009assessing}.

Fuzzy-logic approaches offer another route from qualitative judgments to probability tables. \citet{kumar2024generalized} derive CPT entries from membership functions, fuzzy IF--THEN rules, inference, and defuzzification, demonstrating the procedure on a software-design and development BN evaluated under best- and worst-case software-metric scenarios. The work broadens the available judgment formats, but it does not compare the fuzzy-logic procedure with established semi-automatic pipelines in an operational decision process.

A complementary strategy combines expert judgments with observations. \citet{huang2023reasoning} proposed parameter estimation under expert prior knowledge (PEUK), integrating sample data and expert knowledge with the maximum entropy principle. Their evaluation in rainstorm disaster chains compared the hybrid approach with data-based estimation and direct expert elicitation. Unlike methods that complete CPTs from expert inputs alone, such approaches use observations to inform parameter estimates. This distinction matters when selecting a quantification strategy: WSA and RNM address settings in which experts must specify relationships without a sufficiently representative dataset for estimating the corresponding CPTs.

Across these families, fewer elicited values do not by themselves determine practical suitability. Method choice also depends on the interactions to be represented, the semantics of the node states, the judgments experts can provide, and the traceability of the generated probabilities~\citep{rohmer2020uncertainties,podofillini2023traceable,vagnoli2022updating,blomaard2025burden}.

\subsection{Weighted Sum Algorithm}

The Weighted Sum Algorithm (WSA) asks experts to identify compatible configurations of parent states, assess the child-node distribution for those configurations, and assign relative weights to the parents. It then combines the assessed distributions through weighted sums to populate the remaining CPT rows~\citep{das04generating}. Compatible configurations are combinations for which the parent influences can be judged together without producing a conceptually contradictory situation. The elicitation focuses on configurations that experts can more readily imagine, which reduces both the number and the difficulty of direct probability assessments.

WSA has been tested in software-effort models, where it reduced elicitation relative to manually completed CPTs while keeping useful predictive performance~\citep{baker2010assessing,baker2010evaluating}. Later work proposed cluster-based elicitation to organize compatible configurations, along with a leaky weighted-sum variant for relationships the assessed parents do not fully capture~\citep{benbrahim2022cluster,liu2024leakywsa}. The method still requires direct probability vectors, judgments about compatible parent states, and relative parent influence. What it delivers in practice depends as much on the elicitation interface and facilitation as on the completion algorithm itself.

\subsection{Ranked Nodes Method}

The Ranked Nodes Method (RNM) is designed for ordinal variables. It maps each ordered state to an interval on a unit scale and represents each parent as a continuous ranked variable within its state's interval. A weighted function combines the parent values, and a doubly truncated distribution around the function value allocates probability mass to the child states~\citep{fenton2007using,laitila2016improving}. Weighted-minimum (WMIN), weighted-maximum (WMAX), and weighted-mean expressions encode different relationships between the parents and child. Parent weights control relative influence, while a variance parameter controls dispersion.

RNM can generate a large CPT from a small set of semantic and numerical choices, but those choices materially affect the result. \citet{laitila2018theoretical} found that RNM can approximate many real-world CPT relationships while also showing sensitivity to user-selected settings. Later developments provide more systematic construction guidance, alternative beta-distribution parameterization, support for continuous nodes, and automated calibration~\citep{laitila2022advancing,mascaro2022beta,laitila2022continuous,cunha2025automatic}. RNM is thus best understood as a family of ordinal quantification procedures whose output depends on the selected expression, weights, variance, and calibration process.

\subsection{Empirical comparisons and research gap}

Table~\ref{tab:related-comparisons} summarizes empirical comparisons and secondary studies relevant to the present work. Earlier comparisons identified trade-offs rather than a consistently dominant method. The direct WSA--RNM comparison by \citet{baker2009towards}, for example, concerned software-effort estimation with one domain expert. Rohmer's cross-domain review calls for comparative exercises across network characteristics and application domains, together with practical guidance on method selection~\citep{rohmer2020uncertainties}. More recent work on reducing expert burden likewise emphasizes the need to consider model fidelity and transparency alongside elicitation demands~\citep{blomaard2025burden}. These complementary perspectives motivate evaluation of quantification pipelines within the decision processes they are intended to support.

\begin{table*}[t]
\centering
\caption{Prior work most closely related to the present WSA--RNM comparison.}
\label{tab:related-comparisons}
\small
\begin{adjustbox}{max width=\textwidth}
\begin{tabular}{p{3.0cm}p{3.2cm}p{3.6cm}p{6.0cm}}
\toprule
Study & Methods & Setting & Finding relevant to this study \\
\midrule
\citet{baker2010assessing,baker2010evaluating} & WSA and manually elicited CPTs & Software-effort estimation & WSA reduced elicitation while retaining useful predictive performance relative to the manually quantified models. \\
\citet{baker2009towards} & WSA, RNM, and piecewise interpolation & Software-effort estimation; one domain expert & RNM required less elicitation, whereas WSA produced the most accurate model in that setting. \\
\citet{mkrtchyan2016methods} & RNM and four interpolation-based methods & Two simplified human-reliability BNs & Methods represented interdependencies and uncertainty differently; greater flexibility increased elicitation cost. \\
\citet{laitila2018theoretical} & RNM & Reference CPT relationships from real applications & RNM provided a good average fit across many relationships, but user choices could substantially affect accuracy. \\
\citet{rohmer2020uncertainties} & Secondary review of CPT derivation, uncertainty propagation, and sensitivity analysis & Discrete BNs across application domains & Distinguishes obtaining probabilities from assessing their consequences; calls for broader comparisons and practical method-selection guidance. \\
\citet{kumar2024generalized} & Fuzzy-logic CPT generation & Software-design and development BN; best- and worst-case scenarios & Qualitative rules generated CPTs, but the procedure was not compared with alternative pipelines in an operational setting. \\
\citet{blomaard2025burden} & Review of expert-burden reduction approaches & Expert-driven BN parameterization & Practical assessment should consider elicitation burden together with model fidelity, transparency, and reproducibility. \\
\bottomrule
\end{tabular}
\end{adjustbox}
\end{table*}

Complementing research on CPT generation, sensitivity analysis examines how changes in model parameters affect probabilities of interest. \citet{leonelli2025diameter} use TVD to define the diameter of a CPT as the largest distance between its rows and derive bounds on the effects of model perturbations. \citet{leonelli2023bnmonitor} provide software-supported sensitivity and robustness analyses through the bnmonitor R package. These contributions connect the construction of a quantified BN to evaluation of its inferential behavior, a distinction that also motivates comparing complete output distributions alongside modes and rankings.

Recent approaches based on large language models (LLMs) introduce additional sources of quantification inputs. Some elicit numerical estimates from a language model~\citep{kruger2025chatgpt,nafar2026extracting}, whereas others extract statistical evidence from publications for subsequent mathematical reconstruction of CPTs~\citep{gottal-matthes-2026-verifiable}. These approaches differ in how probability inputs are obtained and checked; they do not remove the need to examine the distributions and decisions produced by the resulting BN. Their implications for expert-supported WSA and RNM quantification are discussed in Section~\ref{sec:discussion}.

In software engineering, BNs have supported defect prediction and expert-based estimation~\citep{fenton2008effectiveness,mendes2012using}, evidence-based and strategic decisions~\citep{misirli2014bayesian,manzano2020method}, process diagnosis, and measurement validity~\citep{perkusich2015procedure,saraiva2020bayesian}. These applications establish a range of uses, while the direct method comparisons summarized in Table~\ref{tab:related-comparisons} provide evidence from software-effort estimation and human-reliability modeling. The present study adds a paired evaluation in two organizational software decision contexts, examining expected model behavior, rankings of recorded alternatives, and complete output distributions. Within each context, the graph, priors, and decision evidence are held fixed. Because WSA and RNM elicit different representations and require different specialist choices, the object of comparison is the implemented pipeline rather than an isolated mathematical formula.

\section{Research method}
\label{sec:method}

This section describes how WSA and RNM were compared within ongoing software decision processes. It introduces the embedded case-study design, organizational setting, VALUE framework and tool, and the two decision contexts. It then reports the participants, model-construction and quantification procedures, and the evaluation and analysis procedures used to answer the research questions.

\subsection{Study design and units of analysis}

We conducted an evaluative embedded case study following established guidance for software engineering case studies~\citep{runeson2009guidelines}. The case comprised one software organization and two embedded decision contexts drawn from its project work. In each context, we compared two quantifications of the multi-parent CPTs in an organization-specific BN. The graph and evidence were fixed: WSA produced one quantification and RNM produced the other. This design enabled a within-context comparison while preserving the organizational meaning of the value factors and decisions. It also allowed the comparison to be interpreted as experience with two complete pipelines in use-oriented model construction, rather than as a benchmark detached from an operational setting.

The study combined three evidence sources: records created during real decision meetings, probability judgments from domain experts, and BN-specialist records produced during model construction. The analysis has two parts. Model walkthroughs examine whether each quantified network produces the expected modal output for specified scenarios (RQ1). Retrospective reconstructions examine the distributions and rankings produced for alternatives discussed in recorded decision meetings (RQ2). The model, not an individual stakeholder or database row, is the object evaluated in each unit.

The study proceeded from existing VALUE records to paired model quantification and then to evaluation. First, the decision records and elicited value factors defined the two contexts and supplied root-node frequencies. Second, domain experts reviewed the graph and supplied the judgments required by WSA and RNM. Third, the resulting models were subjected to the same walkthrough scenarios within each context. Finally, recorded alternatives were entered as evidence in both models and their Overall Value distributions and rankings were compared.

\subsection{Organizational setting}

The study took place at VIRTUS, a university-based software R\&D laboratory in Brazil with approximately 300 collaborators\footnote{https://www.virtus.ufcg.edu.br/}. The organization conducts RDI projects and develops software products and prototypes in partnership with external organizations. Its portfolio includes web and mobile systems, as well as projects in embedded and pervasive computing and finance. The decisions analyzed in this study arose from ongoing project work and were not tasks created solely for the evaluation.

Projects followed a Scrum-based process. A typical development team comprised three to five developers led by a project manager and worked with product representatives from the partner organization. Two cross-project groups supported the teams. The software quality assurance group assessed process and delivery quality and performed independent testing, while the user-experience group designed user interfaces and interaction flows. This setting made feature and interface-design choices recurring, multi-stakeholder decisions rather than isolated modeling exercises.

\subsection{The VALUE framework and Value tool}
\label{subsec:value}

VALUE is a framework for making value considerations explicit in decisions about software-intensive products and services~\citep{mendes2018using,mendes2017towards}. It connects stakeholder-defined value factors, records of actual decisions, and organization-specific Bayesian models through five activities:

\begin{description}
    \item[A1 - Discover value factors.] Stakeholders identify the propositions they use to judge value, such as customer importance, feasibility, effort, or usability. The resulting vocabulary is specific to the organization and decision context.
    \item[A2 - Use the factors in decisions.] During decision meetings, stakeholders use the \emph{Value} web tool to assess how each alternative affects the elicited factors. The tool stores the assessments, alternatives, participants, and resulting selections~\citep{freitas2017value}.
    \item[A3 - Construct a value-estimation model.] The factors and accumulated decision records provide the starting point for an organization-specific BN. Domain knowledge defines and quantifies relationships that cannot be learned reliably from the available records.
    \item[A4 - Evaluate the model.] A Model Walkthrough inspects the model's response to specified evidence, and an Outcome Adequacy assessment compares model outputs with recorded decisions.
    \item[A5 - Use the model in later decisions.] The evaluated BN can be incorporated into the Value tool as an additional source of information during decision meetings.
\end{description}

Uppercase \emph{VALUE} therefore denotes the framework, whereas italicized \emph{Value} denotes its web tool. The present study concerns A3 and A4. Data captured in A1 and A2 were used to determine the factors, root-node frequencies, and historical alternatives for comparing the quantification methods. Earlier studies reported the use of the WSA-quantified models to investigate value-based software engineering~\citep{mendes2018using,mendes2019using}; the RNM quantification was collected in parallel and is analyzed here in direct comparison with WSA.

\subsection{Selection and description of the decision contexts}

The candidate units were drawn from earlier applications of the VALUE framework~\citep{mendes2018using,mendes2019using}, in which the organization had completed activities A1 and A2. We selected contexts that had enough recorded alternatives for both planned evaluation procedures and for which domain experts involved in the original decisions remained available. Two contexts met these criteria. Table~\ref{tab:contexts} summarizes their decisions, participants, and evidence. The database counts are factor-level assessment records, not independent observations.

\begin{table*}[t]
\centering
\caption{Decision contexts, participants, and evidence sources.}
\label{tab:contexts}
\small
\begin{adjustbox}{max width=\textwidth}
\begin{tabular}{p{1.0cm}p{3.2cm}p{3.0cm}p{3.0cm}p{4.4cm}}
\toprule
Unit & Decision & Stakeholders & Domain experts & Data used \\
\midrule
A & Select features for forthcoming sprints in two Android Internet-of-Things projects & Two project managers and one product owner & Two project managers & Sprint 9: 385 factor assessments for 10 features, used to estimate root-node priors. Sprint 10: seven feature alternatives used in the retrospective reconstruction. \\
B & Select user-interface designs for a web project-management product & One product owner, three developers, and one web designer & One project manager and one developer & Meetings 4, 18, and 20: 744 factor assessments for 18 designs, used to estimate root-node priors. Meetings 16 and 17: 293 assessments for nine alternatives used in the retrospective reconstruction. \\
\bottomrule
\end{tabular}
\end{adjustbox}
\end{table*}

\paragraph{Context A: feature selection}
The first context involved two Android Internet-of-Things applications: one for remotely controlling plant watering, and another for monitoring and controlling water level and flow in tanks. Two project managers and one product owner used the Value tool to assess candidate features for Sprints 9 and 10. Their earlier elicitation had produced 13 value factors, and the BN added three intermediate or output constructs (Developer Perspective, Sprint Success, and Overall Value), for a total of 16 nodes. Sprint 9 contained 10 evaluated features and supplied 385 factor assessments for root-node priors. The seven features discussed for Sprint 10 supplied the historical alternatives used for RQ2.

\paragraph{Context B: user-interface design selection}
The second context concerned a web application for project management and continuous improvement in agile teams. A product owner, three developers, and a web designer used the Value tool to assess candidate screen designs. Ten elicited value factors described business, technical, and user-experience concerns; the BN added Effort Reduction, Usability, and Overall Value, producing 13 nodes. Eighteen designs discussed in meetings 4, 18, and 20 supplied 744 factor assessments for the root-node priors. Nine alternatives discussed in meetings 16 and 17 supplied 293 assessments for RQ2. Because the prior calculation included meetings 18 and 20, the reconstruction is descriptive and does not constitute a chronological training--test split.

A subsequent written consultation with a VIRTUS technical lead provided a contemporary perspective on these decision contexts. The respondent had approximately nine years of relevant professional experience, including five at VIRTUS, and authorized publication of the complete responses. The account described ongoing requirements and feature-prioritization activities, together with a recent project involving the modernization of an existing software interface. Decisions involved balancing client expectations, usability, implementation feasibility, and development effort. Large language models (LLMs) were reported to support requirements structuring, initial prioritization, roadmap preparation, and the generation of interface alternatives, while professional review and client validation remained part of the process. This account illustrates the continued relevance of the decision problems examined here, alongside changes in the tools used to address them. The questions and complete responses, translated from Portuguese into English, are provided in \ref{app:practitioner-consultation}.

\subsection{Participants and study preparation}

Two domain experts took part in BN construction for each context. Context A involved two project managers; Context B involved one project manager and one developer. In each context, one expert reported 1--3 years and the other 4--6 years of professional experience in software development-related activities. All four had participated in the relevant VALUE activities and were familiar with the projects and the meaning of the factors. Before model construction, they received a 20-minute introduction covering BN graphs, probabilistic reasoning, and what-if analysis using a simple example.

Three researchers with BN expertise supported quantification: one facilitated WSA and two supported RNM. WSA elicitation was conducted with a Netica plug-in during facilitated sessions. RNM target distributions were collected in spreadsheets and then processed with specialist support. These procedures define the two pipelines compared in the study.

Participation in the elicitation sessions and walkthrough review was voluntary and informed; participants were told how their judgments would be used in this study before contributing. Given the internal, process-improvement nature of these activities within the organization's ongoing project work, formal ethics-committee review was not sought.

\subsection{Model construction and quantification}

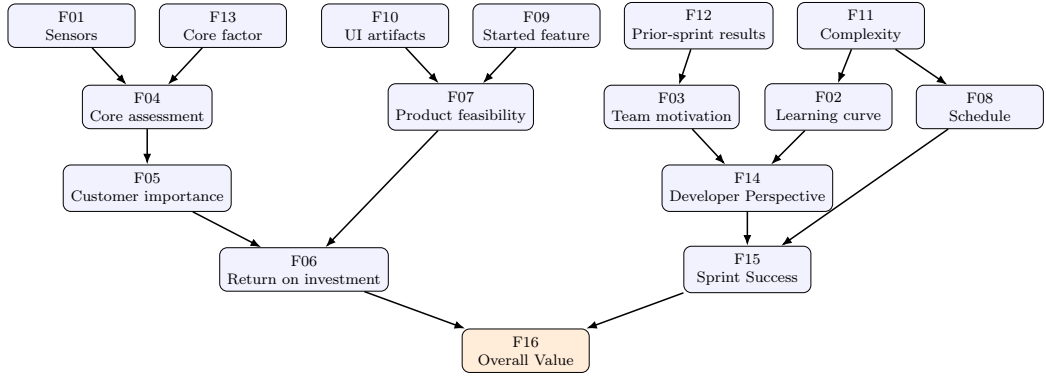
\begin{figure*}[t]
\centering
\resizebox{\textwidth}{!}{%
\begin{tikzpicture}[
  bnnode/.style={draw,rounded corners,align=center,font=\scriptsize,minimum width=2.35cm,minimum height=.8cm,fill=blue!5},
  output/.style={bnnode,fill=orange!15},
  edge/.style={-{Latex[length=2mm]},thick}
]
\node[bnnode] (f1)  at (0,5.0) {F01\\Sensors};
\node[bnnode] (f13) at (2.8,5.0) {F13\\Core factor};
\node[bnnode] (f10) at (5.8,5.0) {F10\\UI artifacts};
\node[bnnode] (f9)  at (8.6,5.0) {F09\\Started feature};
\node[bnnode] (f12) at (11.6,5.0) {F12\\Prior-sprint results};
\node[bnnode] (f11) at (14.6,5.0) {F11\\Complexity};
\node[bnnode] (f4)  at (1.4,3.5) {F04\\Core assessment};
\node[bnnode] (f7)  at (7.2,3.5) {F07\\Product feasibility};
\node[bnnode] (f3)  at (11.1,3.5) {F03\\Team motivation};
\node[bnnode] (f2)  at (14.0,3.5) {F02\\Learning curve};
\node[bnnode] (f8)  at (16.8,3.5) {F08\\Schedule};
\node[bnnode] (f5)  at (1.4,2.0) {F05\\Customer importance};
\node[bnnode] (f14) at (12.5,2.0) {F14\\Developer Perspective};
\node[bnnode] (f6)  at (4.3,.5) {F06\\Return on investment};
\node[bnnode] (f15) at (12.5,.5) {F15\\Sprint Success};
\node[output] (f16) at (8.4,-1.0) {F16\\Overall Value};
\draw[edge] (f1)--(f4); \draw[edge] (f13)--(f4); \draw[edge] (f4)--(f5);
\draw[edge] (f5)--(f6); \draw[edge] (f10)--(f7); \draw[edge] (f9)--(f7); \draw[edge] (f7)--(f6);
\draw[edge] (f12)--(f3); \draw[edge] (f11)--(f2); \draw[edge] (f11)--(f8);
\draw[edge] (f3)--(f14); \draw[edge] (f2)--(f14); \draw[edge] (f14)--(f15); \draw[edge] (f8)--(f15);
\draw[edge] (f6)--(f16); \draw[edge] (f15)--(f16);
\end{tikzpicture}}
\caption{Structure of the Context A Bayesian value-estimation model.}
\label{fig:case-a-bn}
\end{figure*}

\begin{figure*}[t]
\centering
\resizebox{\textwidth}{!}{%
\begin{tikzpicture}[
  bnnode/.style={draw,rounded corners,align=center,font=\scriptsize,minimum width=2.35cm,minimum height=.8cm,fill=blue!5},
  output/.style={bnnode,fill=orange!15},
  edge/.style={-{Latex[length=2mm]},thick}
]
\node[bnnode] (f3) at (0,4.5) {F03\\Third-party libraries};
\node[bnnode] (f2) at (2.8,4.5) {F02\\Complexity};
\node[bnnode] (f8) at (5.6,4.5) {F08\\Clicks};
\node[bnnode] (f6) at (8.4,4.5) {F06\\UI items};
\node[bnnode] (f7) at (11.2,4.5) {F07\\Navigation flow};
\node[bnnode] (f9) at (14.0,4.5) {F09\\Meaningful UI};
\node[bnnode] (f5) at (16.8,4.5) {F05\\Color palette};
\node[bnnode] (f11) at (1.4,2.7) {F11\\Effort Reduction};
\node[bnnode] (f10) at (8.4,2.7) {F10\\UI simplicity};
\node[bnnode] (f12) at (14.0,2.7) {F12\\Usability};
\node[bnnode] (f1) at (3.8,.8) {F01\\UI requirements};
\node[bnnode] (f4) at (13.0,.8) {F04\\Reuse potential};
\node[output] (f13) at (8.4,-1.0) {F13\\Overall Value};
\draw[edge] (f3)--(f11); \draw[edge] (f2)--(f11);
\draw[edge] (f8)--(f10); \draw[edge] (f6)--(f10); \draw[edge] (f7)--(f10);
\draw[edge] (f10)--(f12); \draw[edge] (f9)--(f12); \draw[edge] (f5)--(f12);
\draw[edge] (f11)--(f13); \draw[edge] (f1)--(f13); \draw[edge] (f12)--(f13); \draw[edge] (f4)--(f13);
\end{tikzpicture}}
\caption{Structure of the Context B Bayesian value-estimation model.}
\label{fig:case-b-bn}
\end{figure*}
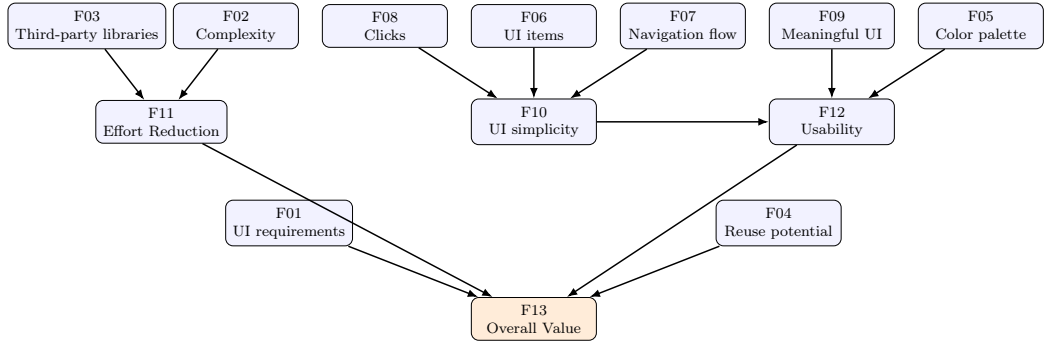

For each context, the domain experts reviewed a graph initialized with the value factors elicited in VALUE activity A1. They clarified factor definitions, added intermediate constructs needed to express the decision logic, and reviewed the directions of the arcs. The graphs were implemented in Netica. Every node was discrete and ordinal, and Overall Value had the states Low, Medium, and High. Table~\ref{tab:factor-definitions} defines the identifiers used in the graphs and parameter table. Figures~\ref{fig:case-a-bn} and~\ref{fig:case-b-bn} show how these nodes are connected in Contexts A and B, respectively.

\begin{table*}[t]
\centering
\caption{Nodes in the two Bayesian value-estimation models.}
\label{tab:factor-definitions}
\scriptsize
\begin{adjustbox}{max width=\textwidth}
\begin{tabular}{clcl}
\toprule
\multicolumn{2}{c}{Context A: feature selection} & \multicolumn{2}{c}{Context B: interface-design selection} \\
\cmidrule(lr){1-2}\cmidrule(lr){3-4}
ID & Factor or model construct & ID & Factor or model construct \\
\midrule
F01 & Sensor implementation & F01 & Satisfaction of UI requirements \\
F02 & Learning curve & F02 & Implementation complexity \\
F03 & Team motivation & F03 & Availability of third-party libraries \\
F04 & Core-functionality assessment (derived) & F04 & Reuse potential \\
F05 & Importance to the customer & F05 & Conformity with the project color palette \\
F06 & Return on investment & F06 & Number of UI items \\
F07 & Feasibility as a product & F07 & Conformity with the standard navigation flow \\
F08 & Schedule compliance & F08 & Number of clicks required \\
F09 & Continuation of an already started feature & F09 & Meaningfulness of the UI \\
F10 & Existing related UI artifacts & F10 & UI simplicity \\
F11 & Feature complexity & F11 & Effort Reduction \\
F12 & Pending results from the previous sprint & F12 & Usability \\
F13 & Core-functionality factor (elicited) & F13 & Overall Value \\
F14 & Developer Perspective & & \\
F15 & Sprint Success & & \\
F16 & Overall Value & & \\
\bottomrule
\end{tabular}
\end{adjustbox}
\end{table*}

Within each context, the WSA and RNM models used the same graph, root-node priors, and manually elicited one-parent CPTs. Only the multi-parent CPTs generated by the two semi-automatic procedures differed. In Context A, frequencies from Sprint 9 populated six root nodes, four one-parent CPTs were elicited directly, and WSA or RNM populated six multi-parent CPTs. For an illustrative two-parent table with three-state parents and a child, WSA reduced the requested probability entries from 27 to 15. In Context B, frequencies from the recorded design assessments populated nine root nodes, and WSA or RNM populated four multi-parent CPTs. For an illustrative table with more parents or higher-cardinality parents, WSA reduced the requested entries from 81 to 27.

\paragraph{WSA procedure}
The facilitator used the Netica plug-in to present each multi-parent relationship. The domain experts identified compatible parent-state configurations, supplied the corresponding child distributions, and stated the relative influence of the parents. The plug-in then generated the unassessed rows using WSA. The experts reviewed the resulting CPTs as part of the modeling session. The recorded time for collecting the WSA inputs for the six multi-parent CPTs in Context A was 93 minutes; for the four multi-parent CPTs in Context B, it was 41 minutes. These observations describe the WSA elicitation sessions only and are not used to compare total effort between WSA and RNM.

\paragraph{RNM procedure}
The domain experts supplied target child distributions for selected combinations of extreme parent states. An RNM specialist chose WMIN or WMAX to represent the direction of the relationship. The \emph{Simulated Specialist} program searched candidate parent weights and variances, using 10,000 simulated values to approximate each candidate output distribution. For $N$ target configurations and $S$ child states, the search minimized the mean squared Euclidean distance per target configuration:

\begin{equation}
L=\frac{1}{N}\sum_{n=1}^{N}\sum_{s=1}^{S}
\left(\widehat{p}_{ns}-p_{ns}\right)^2,
\label{eq:calibration-loss}
\end{equation}

where $p_{ns}$ is the expert-supplied target probability and $\widehat{p}_{ns}$ is the RNM output. The normalization is by configurations rather than by individual probability entries. The candidate with the smallest value of Eq.~\ref{eq:calibration-loss} was retained. Table~\ref{tab:rnm-parameters} reports the resulting parameterizations; integer weights express relative influence.

\begin{table*}[t]
\centering
\caption{RNM parameters used in the two contexts.}
\label{tab:rnm-parameters}
\small
\begin{adjustbox}{max width=\textwidth}
\begin{tabular}{lllrlr}
\toprule
Context & Child & Expression & $\sigma^2$ & Parent weights & Loss $L$ \\
\midrule
A & F16 & WMAX & 0.0400 & F06:4, F15:4 & $5.00\times10^{-3}$ \\
A & F04 & WMAX & 0.0620 & F01:4, F13:4 & $3.94\times10^{-6}$ \\
A & F07 & WMAX & 0.1005 & F10:7, F09:5 & $6.60\times10^{-6}$ \\
A & F06 & WMAX & 0.0635 & F05:7, F07:4 & $1.28\times10^{-4}$ \\
A & F14 & WMAX & 0.0650 & F03:5, F02:7 & $1.09\times10^{-5}$ \\
A & F15 & WMAX & 0.1350 & F14:9, F08:9 & $1.43\times10^{-2}$ \\
\midrule
B & F13 & WMIN & 0.0405 & F01:10, F12:10, F04:1, F11:1 & $2.55\times10^{-2}$ \\
B & F11 & WMAX & 0.0260 & F03:4, F02:10 & $3.75\times10^{-7}$ \\
B & F12 & WMIN & 0.0720 & F10:10, F09:7, F05:5 & $2.09\times10^{-3}$ \\
B & F10 & WMIN & 0.0915 & F08:9, F06:9, F07:10 & $2.47\times10^{-3}$ \\
\bottomrule
\end{tabular}
\end{adjustbox}
\end{table*}

\subsection{Evaluation and analysis}

\subsubsection{RQ1: walkthrough-scenario agreement}

The Model Walkthrough comprised seven scenarios for Context A and eight for Context B. The hypothetical scenarios were prepared by the BN specialists as part of the model-validation activities, while the expected state for Overall Value reflected the domain experts' expected modal outcome for each scenario. Each scenario specified states for a subset of input factors. We entered the same evidence into the WSA and RNM models and recorded the modal output state. Context A's walkthrough forms called the three expected outcomes Positive, Neutral, and Negative; Table~\ref{tab:scenario-results} reports the corresponding Overall Value labels High, Medium, and Low. For each model, we counted exact matches with the expected label and identified the scenarios on which the outputs differed.

The expected labels represent expert-defined model-review judgments rather than independently observed outcomes, and the scenarios were not prepared as an independent test set for either quantification pipeline. They therefore assess whether each quantified model responds consistently with the domain expectations documented during model construction. We report scenario-level results and agreement counts without an inferential test.

\subsubsection{RQ2: retrospective selection concordance}

For Context A, the factor assessments for the seven features discussed for Sprint 10 were converted into BN evidence. After d-separation, features 2--5 produced the same effective evidence, leaving four distinguishable patterns. For Context B, the nine designs discussed in meetings 16 and 17 were reconstructed from their recorded assessments. Designs 2 and 7 produced one common pattern and designs 6 and 9 another. In addition, observing F10 made F6--F8 irrelevant to Overall Value, leaving seven distinguishable patterns.

For every alternative, we recorded the complete Overall Value distribution and ranked alternatives by $\phigh$. We then compared the ranking with the recorded selected/not-selected status. To describe distributional differences between the pipelines, we calculated total variation distance for each distinguishable evidence pattern:

\begin{equation}
\operatorname{TVD}(p^{\mathrm{WSA}},p^{\mathrm{RNM}})
=\frac{1}{2}\sum_{s\in\{\mathrm{Low,Medium,High}\}}
\left|p^{\mathrm{WSA}}_s-p^{\mathrm{RNM}}_s\right|.
\label{eq:tvd}
\end{equation}

The TVD in Eq.~\ref{eq:tvd} ranges from 0 for identical distributions to 1 for distributions with no overlapping mass. We report the range and median within each context. Historical selection is not treated as an objective value label: stakeholder choices can incorporate dependencies, capacity, strategy, and negotiation that are outside the BN. The analysis therefore concerns ranking concordance and distributional differences, not classification or prospective predictive accuracy.

\section{Results}
\label{sec:results}

This section reports the findings for the two research questions. It first
presents the agreement between each quantified model and the predefined
walkthrough expectations (RQ1). It then reports retrospective selection
concordance, rankings, and distribution-level differences between WSA and RNM
(RQ2).

\subsection{RQ1: WSA agreed with more walkthrough labels in both contexts}

Table~\ref{tab:scenario-results} reports every modal state. In Context A, WSA matched all seven expected labels; RNM matched four, differing on scenarios 2, 4, and 7. In Context B, WSA matched six of eight labels and RNM matched four. Both models differed from the expectation on scenarios 2 and 3; RNM also differed on scenarios 1 and 8. Across the two walkthroughs, WSA more often reproduced the domain expectations encoded in the scenarios.

\begin{table*}[t]
\centering
\caption{Modal-state agreement in the walkthrough scenarios. Bold outputs differ from the expected label.}
\label{tab:scenario-results}
\small
\begin{tabular}{cllll}
\toprule
Context & Scenario & Expected label & WSA output & RNM output \\
\midrule
A & 1 & High & High & High \\
A & 2 & Low & Low & \textbf{Medium} \\
A & 3 & High & High & High \\
A & 4 & Low & Low & \textbf{High} \\
A & 5 & High & High & High \\
A & 6 & High & High & High \\
A & 7 & Low & Low & \textbf{High} \\
\cmidrule{2-5}
  & Agreement & & 7/7 & 4/7 \\
\midrule
B & 1 & Low & Low & \textbf{Medium} \\
B & 2 & Low & \textbf{High} & \textbf{Medium} \\
B & 3 & Low & \textbf{Medium} & \textbf{Medium} \\
B & 4 & Low & Low & Low \\
B & 5 & High & High & High \\
B & 6 & Low & Low & Low \\
B & 7 & High & High & High \\
B & 8 & Low & Low & \textbf{Medium} \\
\cmidrule{2-5}
  & Agreement & & 6/8 & 4/8 \\
\bottomrule
\end{tabular}
\end{table*}

\subsection{RQ2: both pipelines placed selected alternatives near the top but produced different distributions}

\paragraph{Context A}
Table~\ref{tab:historical-a} shows the seven feature alternatives. WSA ranked selected features 7, 1, and 6 as its three largest values of $\phigh$; RNM placed selected features 1 and 7 first (tied), followed by selected feature 6. Feature 3 was selected but tied under both methods with rejected features 2, 4, and 5 because those alternatives reduce to the same effective evidence pattern. The models therefore placed selected alternatives at the top but did not distinguish every selected feature from every rejected one.

RNM assigned more mass to High than WSA for every feature in this set and assigned no simulated mass to Low at the reported precision. Across the four distinguishable evidence patterns, TVD ranged from 0.250 to 0.380 (median 0.304). The pipelines had the same modal state for all four patterns, although their probability allocations differed.

\begin{table*}[t]
\centering
\caption{Overall Value distributions for the Context A retrospective reconstruction. RNM percentages are Monte Carlo proportions; displayed 0.0 and 100.0 values are rounded rather than analytical point masses.}
\label{tab:historical-a}
\small
\begin{tabular}{crrrcrrr}
\toprule
& \multicolumn{3}{c}{WSA (\%)} && \multicolumn{3}{c}{RNM (\%)} \\
\cmidrule(lr){2-4}\cmidrule(lr){6-8}
Feature & Low & Medium & High & Selected & Low & Medium & High \\
\midrule
1 & 17.3 & 13.4 & 69.3 & Yes & 0.0 & 0.0 & 100.0 \\
2 & 38.0 & 20.0 & 42.0 & No & 0.0 & 39.0 & 61.0 \\
3 & 38.0 & 20.0 & 42.0 & Yes & 0.0 & 39.0 & 61.0 \\
4 & 38.0 & 20.0 & 42.0 & No & 0.0 & 39.0 & 61.0 \\
5 & 38.0 & 20.0 & 42.0 & No & 0.0 & 39.0 & 61.0 \\
6 & 25.0 & 19.0 & 56.0 & Yes & 0.0 & 25.0 & 75.0 \\
7 & 17.0 & 13.0 & 70.0 & Yes & 0.0 & 0.0 & 100.0 \\
\bottomrule
\end{tabular}
\end{table*}

\paragraph{Context B}
Table~\ref{tab:historical-b} shows the nine design alternatives. Designs 2 and 7 were selected. Under both pipelines they tied for the largest $\phigh$ (WSA: 76.0\%; RNM: 94.0\%). Several rejected designs also had High as their modal state, including designs 1, 4, 5, and 8 under RNM and designs 4--6 and 8--9 under WSA. The models identified the selected designs as the strongest candidates but did not provide a selected/not-selected separation.

The distributions differed most visibly for design 1 (WSA modal Low at 77.0\%, RNM modal High at 89.0\%), design 3 (WSA modal Low at 77.0\%, RNM Medium at 100.0\%), and design 6 (WSA modal High at 70.0\%, RNM Medium at 95.0\%). Across the seven distinguishable evidence patterns, TVD ranged from 0.210 to 0.890 (median 0.340), and the modal states differed for three patterns.

\begin{table*}[t]
\centering
\caption{Overall Value distributions for the Context B retrospective reconstruction. RNM percentages are Monte Carlo proportions; displayed 0.0 and 100.0 values are rounded.}
\label{tab:historical-b}
\small
\begin{tabular}{crrrcrrr}
\toprule
& \multicolumn{3}{c}{WSA (\%)} && \multicolumn{3}{c}{RNM (\%)} \\
\cmidrule(lr){2-4}\cmidrule(lr){6-8}
Design & Low & Medium & High & Selected & Low & Medium & High \\
\midrule
1 & 77.0 & 3.0 & 20.0 & No & 0.0 & 11.0 & 89.0 \\
2 & 21.0 & 3.0 & 76.0 & Yes & 0.0 & 6.0 & 94.0 \\
3 & 77.0 & 14.0 & 9.0 & No & 0.0 & 100.0 & 0.0 \\
4 & 3.0 & 39.0 & 58.0 & No & 0.0 & 11.0 & 89.0 \\
5 & 34.0 & 31.5 & 34.5 & No & 0.0 & 46.0 & 54.0 \\
6 & 24.0 & 6.0 & 70.0 & No & 0.0 & 95.0 & 5.0 \\
7 & 21.0 & 3.0 & 76.0 & Yes & 0.0 & 6.0 & 94.0 \\
8 & 24.0 & 4.0 & 72.0 & No & 0.0 & 11.0 & 89.0 \\
9 & 24.0 & 6.0 & 70.0 & No & 0.0 & 95.0 & 5.0 \\
\bottomrule
\end{tabular}
\end{table*}

\section{Discussion}
\label{sec:discussion}

This section interprets the findings and their implications: first what the walkthrough, ranking, and distribution-level results mean in light of previous evidence, then what they imply for software decision-support practice and for research on BN quantification.

\subsection{Interpretation of the findings}

The two evaluation procedures probe different things. The walkthrough checks whether the completed CPTs propagate specified evidence in line with the qualitative expectations used during model review, and WSA reproduced more of those expectations in both contexts. The retrospective reconstruction instead checks whether the model's rankings match the choices made in earlier meetings. Both methods placed the selected candidates near the top, including an exact top tie for the two selected designs in Context B, but neither reproduced all selected/not-selected distinctions.

The full distributions add information the modes and rankings hide. In Context A, the methods agreed on the modal state for all four distinct evidence patterns, while TVD still reached 0.380. In Context B, TVD reached 0.890, and three of seven patterns had different modes. If a BN is used only to produce a shortlist, the leading common candidates look like a useful agreement. Once the probabilities feed into later inference, portfolio analysis, or expected-utility calculations, the distributional differences stop being cosmetic.

RNM placed little or no simulated mass on Low for the reconstructed alternatives. This traces back to the complete RNM pipeline used here: the elicited extreme-state targets, the selected WMIN/WMAX expressions, the parent weights, the variance, and the calibration search jointly determine the CPTs. The study compares the pipeline with the implemented WSA pipeline; it does not isolate the effect of any single RNM component.

The calibration loss $L$ reported in Table~\ref{tab:rnm-parameters} offers a partial, mechanistic explanation for where the largest distributional divergence shows up. In Context B, the Overall Value node (F13) carries the highest calibration loss in that context ($L=2.55\times10^{-2}$), an order of magnitude above most other Context B nodes. That same node's CPT directly generates the distributions used for TVD comparisons, and Context B correspondingly shows both a higher median TVD and the most extreme value (0.890). Context A's Overall Value node (F16), by contrast, was calibrated with a lower loss ($L=5.00\times10^{-3}$). We treat this as a plausible contributing factor, not an isolated causal test, since the calibration search, elicited targets, and expression choice were not varied independently.

\subsection{Relationship to previous evidence}

Earlier reports from these decision contexts investigated value-based software engineering and the VALUE framework using only WSA-quantified models~\citep{mendes2018using,mendes2019using}. The paired reconstruction reported here adds a different kind of evidence. RNM also placed the alternatives selected by stakeholders near the top, so retrospective ranking concordance was not unique to WSA. At the same time, the two pipelines sometimes represented the uncertainty behind those rankings in very different ways. This complements the earlier organizational findings: a Bayesian value-estimation model may yield the same immediate prioritization outcome across multiple quantification pipelines, but the choice of pipeline can still matter once complete probabilities are reused for later inference or decision analysis.

The higher WSA walkthrough agreement fits the broad pattern reported by \citet{baker2009towards}: higher accuracy for WSA and lower elicitation effort for RNM in a software-effort model. The outcome definitions differ substantially, though. Our walkthrough measures agreement with scenario expectations, not prediction error relative to observed software effort, so the two studies point in the same direction regarding WSA's ability to preserve elicited judgments without providing a pooled estimate of comparative accuracy.

The observed differences between the WSA and RNM distributions highlight the practical relevance of RNM parameter choices. \citet{laitila2018theoretical} showed that RNM can fit many relationships while remaining sensitive to user-selected settings. Our parameter search operationalized those choices, and Table~\ref{tab:rnm-parameters} reports the retained parameterizations. The Context B results show that the calibrated RNM pipeline and WSA can still allocate probability mass differently. This finding is consistent with Rohmer's distinction between deriving CPTs and evaluating the consequences of their assumptions~\citep{rohmer2020uncertainties}. Together with earlier comparative and review evidence, it reinforces the importance of matching elicitation burden, flexibility, and traceability to the modeled relationship and its implementation~\citep{mkrtchyan2016methods,blomaard2025burden}.

The use of TVD also connects our findings to recent work on BN sensitivity. \citet{leonelli2025diameter} use distances between rows of a CPT to characterize dependence and bound the effects of perturbations. Here, TVD measures a different object: the difference between output distributions produced by WSA and RNM under the same evidence. The two perspectives are complementary. Our results identify decision scenarios in which the implemented pipelines yield different probability allocations despite similar rankings; sensitivity analysis can then investigate which parameter changes affect probabilities or decisions of interest. In particular, similar shortlists do not establish that conclusions based on probability thresholds or expected utilities would also remain unchanged.

Recent LLM-based parameterization studies extend the comparison from how probabilities are completed to how the initial numerical judgments are obtained. \citet{kruger2025chatgpt} used ChatGPT to parameterize two BN classifiers for air surveillance and assessed both class assignments and probability outputs against an expert-parameterized reference network. Providing additional operational context improved the reported performance, while more complex CPTs required corrections to their structure and normalization. In a broader evaluation of 80 BNs, \citet{nafar2026extracting} found that LLM-generated probability estimates could provide informative priors, with benefits from combining them with data particularly in low-data settings. Their evaluation also examined downstream classification, connecting parameter quality to the eventual use of the network.

A different approach separates evidence extraction from numerical reconstruction. \citet{gottal-matthes-2026-verifiable} used LLMs to extract statistical summaries from scientific publications and reconstructed CPTs through explicit mathematical procedures. They evaluated parameter fidelity, posterior inference, and decision utility separately. This distinction is relevant to the present findings: the similar placement of selected alternatives under WSA and RNM did not imply similar probability distributions. The comparison therefore identifies an evaluation requirement that also matters for LLM-assisted quantification: agreement on a leading alternative should be assessed alongside the uncertainty represented by the complete distribution and its consequences for subsequent inference.

\subsection{Implications for software decision-support practice}

The case suggests a practical sequence for organizations considering semi-automatic CPT construction. First, pick a pipeline whose assumptions align with the modeled nodes and whose questions match the judgments that domain experts can actually provide. WSA works when experts can assess representative probability distributions and relative parent influence. RNM is attractive for ordinal nodes when the organization can justify the ranked expression, weights, variance, and calibration targets. Familiarity with an elicitation format can shape this choice, but it does not replace the need to validate the generated CPTs.

Second, evaluate the model at whatever level its intended use actually demands. Walkthrough scenarios check whether the model reacts consistently with the causal and value assumptions used during construction. Rankings matter directly when the BN supports shortlisting or prioritization. Complete distributions need separate inspection when probabilities will be propagated, combined with utilities, or reused in later analyses. This case shows why these checks are not interchangeable: both pipelines identified strong candidates, yet some probability allocations differed substantially.

Third, keep an audit trail for the complete pipeline: the graph version, priors, elicited distributions, parent weights, RNM expressions and calibration targets, software settings, and judgments revised during review. A small pilot on representative multi-parent CPTs can reveal whether experts understand the requested judgments before the whole network is quantified. For consequential decisions, examine whether plausible changes to those inputs alter either the ranking of alternatives or the probability magnitudes used in the decision rule. Rohmer's review motivates this separation between elicitation and evaluation, while tools such as bnmonitor provide procedures for examining the sensitivity of probabilities of interest~\citep{rohmer2020uncertainties,leonelli2023bnmonitor}.

LLM assistance can also be organized around expert review of evidence-grounded suggestions. Focusing on network structure and clinician interaction, \citet{cypko2026large} evaluated a BN modeling tool combining LLMs with retrieval-augmented generation (RAG) with four clinicians. The study illustrates a workflow in which users inspect, accept, or revise suggestions based on retrieved domain knowledge. For software decision-support models, a corresponding opportunity is to ground proposed quantification inputs in project records and explicit definitions of the organization's value factors. Such assistance could help prepare judgments for WSA or RNM while leaving their contextual interpretation and acceptance with the responsible stakeholders.

\subsection{Implications for research on BN quantification}

The evidence comes from two software decision-support models, but the comparison points to evaluation dimensions that apply more broadly to expert-driven ordinal BNs: the form of elicited judgments, facilitator involvement, calibration procedure, and the gap between modal and distribution-level agreement. Empirical comparisons should treat a quantification method as an elicitation pipeline rather than a single algorithm. A controlled follow-up could equalize facilitator contact and information budgets, record time separately for domain experts and BN specialists, and vary the completion method while holding everything else constant.

Prospective evaluation against decisions or outcomes not used during model construction would complement the walkthrough and retrospective procedures used here. A further step is to connect differences between quantification pipelines to the sensitivity of their downstream outputs. For WSA, this could involve varying representative distributions and parent weights; for RNM, expressions, weights, variances, target vectors, and numerical settings warrant investigation. Automated RNM calibration offers a useful comparison point~\citep{cunha2025automatic}, while sensitivity methods implemented in bnmonitor provide ways to examine how parameter perturbations affect target probabilities~\citep{leonelli2023bnmonitor}. Such analyses could distinguish distributional differences that leave the intended decision unchanged from those that change a ranking, cross a decision threshold, or alter an expected-utility comparison.

Building on recent LLM-based parameterization and evidence-extraction approaches~\citep{nafar2026extracting,gottal-matthes-2026-verifiable}, future studies could investigate assistance with the representative distributions elicited for WSA and the extreme-configuration targets used to calibrate RNM. An LLM could propose candidate distributions using descriptions of the nodes, their states, and relevant organizational evidence, which experts would then inspect and revise before CPT generation. Comparing each pipeline with and without this assistance, while holding node definitions and elicitation tasks constant, would help establish whether it improves the elicitation process. Evaluation should cover complete distributions, downstream rankings or decisions, and sensitivity to model and prompt choices. It should also measure total expert effort, including review and correction, together with perceived workload and computational cost. Independent evaluation scenarios would distinguish agreement with the inputs used during generation or calibration from performance on new decision situations.

Replication across organizations, contemporary decision records, and decision types is still needed before comparative performance can be generalized. In the meantime, future studies can treat the three evaluation levels used here (expected model behavior, decision ranking, and complete-distribution comparison) as separate outcomes worth reporting on their own, rather than as stand-ins for one another.

\section{Threats to validity and study limitations}
\label{sec:threats}

We organize the threats to validity using the construct, internal, conclusion, and external validity categories from \citet{wohlin2012experimentation}, and we also discuss reliability and reproducibility, since the comparison depends on elicitation records, parameterizations, and software-supported quantification procedures.

\paragraph{Construct validity}
``Overall Value'' is a company-specific, three-state ordinal construct. The walkthrough labels represent internally developed expectations, not observed outcomes. Historical selection is also an imperfect proxy for value, since decisions may reflect dependencies, capacity, strategy, or negotiation outside the BN. RQ2 is accordingly restricted to selection concordance, with the complete probability vectors analyzed alongside modal agreement.

\paragraph{Internal validity}
The pipelines were operationalized differently. WSA was elicited with an in-person BN facilitator, while RNM inputs were collected remotely in spreadsheets and processed by two RNM specialists. As a result, interaction mode, facilitator allocation, tools, and calibration are part of the comparison. In Context B, priors incorporated assessments from meetings 18 and 20, while some reconstructed alternatives came from meetings 16 and 17. The retrospective results should be read as model reconstructions, not temporally held-out performance.

\paragraph{Conclusion validity}
There is one organization, two purposefully selected contexts, 15 walkthrough scenarios, seven feature alternatives (four distinguishable evidence patterns), and nine design alternatives (seven distinguishable patterns). Alternatives share participants, model structure, priors, and meetings, and are not independent replications. The analysis is descriptive and within-context; the agreement counts, and TVD summaries should not be read as population estimates.

\paragraph{External validity}
The value factors, graphs, participants, and calibration targets are organization-specific. Both contexts concern value-based prioritization and were chosen because sufficient records and knowledgeable participants were available. The original decision records reflect the practices of those earlier projects. The subsequent consultation illustrates the continued relevance of feature and interface decisions, alongside changes such as LLM-assisted requirements and design work. This contextual account from one professional does not establish that the original value factors or comparative WSA--RNM performance transfer unchanged to current workflows. Changes in tools and processes may also affect elicitation effort. Generalization to contemporary settings, other organizations, BN topologies, state cardinalities, and decision types therefore requires further empirical evaluation.

\paragraph{Reliability and reproducibility}
The graphs, factor definitions, retained RNM parameterizations, evaluation procedure, and model outputs are reported in the article. This supports structural and analytical traceability, but detailed organizational records and WSA elicitation artifacts cannot be released, and the complete RNM calibration search settings are not reported. We did not analyze sensitivity to alternative RNM calibration settings. Time was not recorded on a commensurate basis for domain experts, facilitators, and calibration work, so the results do not compare the total effort of the pipelines. In addition, the related-work search was targeted rather than systematic, and relevant studies may therefore have been missed.

\section{Conclusion}
\label{sec:conclusion}

Expert-driven Bayesian networks give software organizations a way to combine sparse records with professional judgment, but the value of a model depends on how its CPTs are quantified. We compared WSA and RNM as complete elicitation and quantification pipelines in two recurring decision contexts within a single software R\&D organization: feature selection and user interface design selection. The case ties method behavior to model walkthroughs and to alternatives recorded in actual decision meetings.

The comparison produced different conclusions at different evaluation levels. WSA reproduced more expert-defined walkthrough expectations than RNM (7/7 versus 4/7 in Context A and 6/8 versus 4/8 in Context B). In the retrospective reconstructions, both pipelines placed the alternatives selected by stakeholders near the top. Similar modes and rankings did not imply similar probability distributions, though: median TVD was 0.304 in Context A and 0.340 in Context B, rising to 0.890 for one design pattern. Two pipelines, in other words, can support a similar shortlist while representing the uncertainty behind that shortlist quite differently.

For software decision-support practice, the results favor a staged selection and validation process over a universal choice between WSA and RNM. Practitioners should first match the method to node semantics and to the judgments experts can actually provide: direct representative distributions and relative influence for WSA, or ordinal relationships and defensible calibration parameters for RNM. They should then check the expected model behavior, decision rankings, and complete distributions based on how the BN will actually be used. Recording all elicited inputs, calibration choices, and software settings matters for later review and maintenance. Familiarity with a method can help with adoption, but it is no substitute for these checks.

The evidence is bounded by one organization, two purposefully selected ordinal models, expert-defined walkthrough expectations, retrospective decision records, and differently implemented pipelines. The study neither compared total elicitation effort nor analyzed sensitivity to alternative RNM calibration settings. Replications should evaluate prospectively observed decisions, balance facilitation across methods, measure expert and specialist effort separately, and test whether distributional differences change downstream inferences or utilities. Such studies would show which lessons transfer across organizations and when pipeline choice actually matters for software decisions.

\section*{Acknowledgments}

The authors thank the domain experts and members of the participating organization who contributed to model construction and to the decision records analyzed in this study.

\section*{Declaration of competing interest}

The authors declare that they have no known competing financial interests or personal relationships that could have appeared to influence the work reported in this paper.

\section*{Data availability}

Public repositories provide software artifacts associated with the RNM and WSA components evaluated in this study:

\begin{itemize}
    \item \href{https://github.com/joaonunes-copin/kaizen-prototype}{RNM prototype repository}
    \item \href{https://github.com/joaonunes-copin/KaizenRNM}{RNM algorithm repository}
    \item \href{https://github.com/joaonunes-copin/wsa-algorithm}{WSA algorithm repository}
\end{itemize}

The article reports the model graphs, factor definitions, RNM parameterizations, and analyzed output distributions. The repositories do not contain the complete organizational data or an executable replication package for the case study. The underlying organizational decision records and detailed elicitation artifacts are confidential and are not publicly available.

\section*{Declaration of generative AI and AI-assisted technologies in the manuscript preparation process}

During the preparation of this work, the authors used OpenAI Codex to support language refinement and manuscript editing. The authors reviewed and edited all AI-assisted output and take full responsibility for the content of the published article.

\appendix
\section[Practitioner consultation on current software decision contexts]{Practitioner consultation on current software decision contexts}
\label{app:practitioner-consultation}

This appendix presents the six questions and the complete written responses from the VIRTUS technical lead consulted about current feature-prioritization and interface-design practices. The questions and responses have been translated from Portuguese into English. The respondent authorized publication of the complete responses.

\subsection[Experience with these decisions]{What is your experience with these decisions?}

I currently work as a technical lead at VIRTUS, where I participate in projects involving requirements elicitation and prioritization, feature definition, and decisions related to product interfaces. I have worked on activities of this kind for approximately nine years, including the last five years at VIRTUS.

\subsection[Presence of these decisions in current projects]{Are these two types of decisions present in current projects?}

Yes. One recent example was a project carried out over the past year that, in addition to its research objectives, sought to improve usability and response time and modernize the interface of a system already used by the client. The existing workflow was complex for end users and had performance problems due to the number of devices involved.

In this project, discussions among the technical team, the design team, and the client were held to define new components and interface alternatives. The client provided initial references showing what they expected, and the designer used these to develop proposals. As the technical lead, I participated in evaluating these proposals, mainly to balance the client's expectations, implementation feasibility, and usability. In some cases, for example, a more elaborate visual solution could introduce unnecessary complexity for the user.

\subsection[Information supporting these choices]{What information supports these choices?}

The decisions were primarily supported by the needs presented by the client, the objectives defined for each screen, feedback obtained during interactions with users, and the assessments of the technical and design teams.

The process usually began with a reference or need presented by the client. The designer produced a prototype, which was initially reviewed internally by me and the other team members involved. Once we arrived at a proposal we considered suitable, it was presented to the client for validation before development. This way, we avoided investing implementation effort in a solution that was not yet aligned with the user's expectations.

\subsection[Handling uncertainty]{How is uncertainty handled when it arises?}

When the available information was insufficient to determine the best alternative, the experience of the professionals involved played an important role. The technical and design teams discussed the proposals, considering aspects such as usability, complexity, and development effort.

When there were disagreements, we sought to agree on a proposal and present it to the client. Client validation was an important step in the process, particularly before starting development. Thus, even when there were different interpretations within the team, the decision was refined through discussion and subsequently validated with the person representing the product's needs.

\subsection[Changes in recent years]{What has changed in recent years?}

One of the main changes has been the adoption of LLMs in the process of defining requirements, features, and interfaces. Previously, turning an initial project description into a roadmap, identifying features and epics, and defining a coherent implementation sequence that respected dependencies among features required considerably more effort.

Today, LLMs can generate an initial structure of features, suggest an implementation order, and support the initial preparation of the roadmap. This proposal is still reviewed by the team and subsequently validated with the client, but it reduces the work needed to reach a first version.

A similar change has occurred in interface development. Previously, reproducing a prototype created by the design team could require substantial implementation effort, which directly influenced discussions about which components should be developed. Today, LLM-based tools have greatly reduced this effort and allow alternatives to be tried out and implemented more quickly. What remains similar is the need for professional assessment and client validation.

\subsection[Use of AI tools in these decisions]{Do AI tools, including LLMs, participate in these decisions?}

Yes. We currently use LLMs to support activities such as generating interface alternatives, structuring requirements, defining an initial set of features, and suggesting priorities. This is done through an internal tool that makes LLMs available within the institutional environment, taking into account data protection requirements, confidentiality, and aspects related to Brazil's General Data Protection Law (LGPD). This allows client information, project data, and research content to be used in these activities without having to send them to services outside the institution.

LLMs are used to support analysis and the generation of alternatives, but they are not responsible for the final decision. The professionals remain responsible for reviewing the suggestions, assessing their suitability for the project context, and validating decisions with the product owner (PO) or the client, when applicable. For interfaces and requirements, AI can support preliminary generation and analysis, while final validation remains a human responsibility.

\bibliography{refs}

\end{document}